\documentclass[12pt]{article}
\pdfoutput=1
\usepackage{draft,amsthm}

\newtheorem{definition}{Definition}
\newtheorem{lemma}{Lemma}
\newtheorem{proposition}{Proposition}
\newtheorem{corollary}{Corollary}
\newtheorem{conjecture}{Conjecture}

\renewcommand{\H}{\text{H}}
\renewcommand{\D}{\text{D}}
\renewcommand{\S}{\text{S}}
\newcommand{\Mon}{\mathbb{M}}
\newcommand{\cl}{\text{Cl}}
\newcommand{\cen}{\text{C}}
\newcommand{\comm}[2]{#1#2=#2#1}

\newcommand{\bN}{\mathbb{N}}
\newcommand{\mygcd}[1]{\gcd(#1)}
\newcommand{\dv}{\,|\,}
\newcommand{\bdv}{\,\bigg|\,}

\newcommand{\T}{\mathcal{T}}
\newcommand{\M}{{V^\natural}}
\newcommand{\Sch}{V}
\newcommand{\Zconst}{Z(\tau)|_{q^0}}

\begin{document}

\begin{titlepage}

\begin{center}

\hfill \\
\hfill \\
\vskip 1cm

\title{Topological modularity of Monstrous Moonshine}

\author{Ying-Hsuan Lin$^{a}$}

\address{${}^a$Jefferson Physical Laboratory, Harvard University, Cambridge, MA 02138, USA}

\email{yhlin@fas.harvard.edu}

\end{center}

\vfill

\begin{abstract}
    We explore connections among Monstrous Moonshine, orbifolds, the Kitaev chain and topological modular forms.
    Symmetric orbifolds of the Monster CFT, together with further orbifolds by subgroups of Monster, are studied and found to satisfy the divisibility property, which was recently used to rule out extremal holomorphic conformal field theories.  
    For orbifolds by cyclic subgroups of Monster, we arrive at divisibility properties involving the full McKay--Thompson series.
    Orbifolds by non-abelian subgroups of Monster are further considered by utilizing the data of Generalized Moonshine.
\end{abstract}

\vfill

\end{titlepage}

\tableofcontents

\section{Introduction and summary}

Physical observables can exhibit automorphic properties due to symmetries and dualities.  
A classic example is the high/low temperature duality of conformally invariant systems in two spacetime dimensions: The partition function $Z(\tau)$ on a torus of complex modulus $\tau$ must be invariant under $\tau \to \tau + 1$ and $\tau \to -1/\tau$ up to overall factors controlled by quantum anomalies.  The first transformation corresponds to a Dehn twist, whereas the second corresponds to exchanging space with (Euclidean) time and rescaling the overall size of the torus by conformal invariance.
Functions $Z(\tau)$ enjoying these automorphic properties are called (elliptic) modular forms; hence, two-dimensional conformally invariant systems are said to exhibit ``modularity''.  In addition, twisted partition functions \cite{Witten:1982df,Witten:1986bf} of supersymmetric quantum field theories (SQFT) exhibit modularity even if not conformal.

Mathematicians have refined the ring of modular forms in many ways, one of which is to lift it to a generalized cohomology theory known as topological modular forms (TMF) \cite{douglas2014topological}.  It has been conjectured \cite{Segal,ST1,ST2} that this refinement captures the \emph{full} set of rigid (invariant under deformations) physical observables of minimally-supersymmetric quantum field theories in two spacetime dimensions.
In this sense, such SQFTs exhibit ``topological modularity'' (some more details are given below).
By now, there is overwhelming evidence \cite{Gaiotto:2018ypj,GPPV,Gaiotto:2019asa,Gaiotto:2019gef,Tachikawa:2021mvw,Tachikawa:2021mby,Theo} supporting the conjecture, but the physical reasoning remains to be understood.

In \cite{Lin:2021bcp}, it was argued by invoking topological modularity that the constant term in the Fourier expansion of the torus partition function $Z(\tau)$ of any bosonic (non-spin) holomorphic conformal field theory (CFT) with central charge $c = 24n$ must satisfy the divisibility property
\ie\label{Divisibility}
    \frac{12}{\mygcd{12,n}} \bdv \Zconst,
\fe 
where
\ie
    q := e^{2\pi i \tau}
\fe 
throughout this note.
The significance of this property is at least twofold:
\begin{enumerate}
    \item Physically \eqref{Divisibility} rules out the existence of extremal holomorphic CFTs \cite{hohn2007selbstduale,hohn2008conformal} for a large set of central charges \cite{Lin:2021bcp}.  
    This family of theories, generalizing the Monster CFT \cite{frenkel1984natural,frenkel1988vertex}, was conjectured by Witten \cite{W} to be the holographic dual of a three-dimensional theory of pure quantum gravity.  
    Nevertheless, before \cite{Lin:2021bcp}, their actual existence was anyone's guess; even at $c=48$, a construction was never found, nor was convincing counterevidence presented.
    \item Mathematically \eqref{Divisibility} is a corollary of the Stolz--Teichner conjecture \cite{Segal,ST1,ST2}, which states that every (0,1) SQFT in two spacetime dimensions can be associated with a class in TMF \cite{douglas2014topological}, and the class is invariant under continuous deformations, including but not limited to renormalization group flows.  See \cite{Gaiotto:2018ypj,GPPV,Gaiotto:2019asa,Gaiotto:2019gef,Tachikawa:2021mvw,Tachikawa:2021mby,Theo} for studies and applications of TMF in physics.
    The non-surjectivity of the map from TMF to the partition function is the origin of \eqref{Divisibility}.
    Holomorphic CFTs provide a simple setting for testing the Stolz--Teichner conjecture and for understanding the divisibility property.\footnote{
        Holomorphic CFTs are tautologically (0,1) SQFTs with a trivial supersymmetric sector.
    }
    For instance, as explained in \cite{Lin:2021bcp}, any holomorphic CFT based on an even self-dual lattice has $\Zconst$ divisible by 24, according to a theorem of Borcherds \cite{borcherds1995automorphic}.
    Proving \eqref{Divisibility} for increasingly more complex families of theories may provide new insight into the nature of the divisibility property.
\end{enumerate}

Given a $c=24$ holomorphic CFT $\T$ with finite group global symmetry $G$, infinite families of holomorphic CFTs can be constructed as tensor products $\T^{\otimes n}$ and their orbifolds \cite{Dixon:1985jw,Dixon:1986jc}.  Since the product theory $\T^{\otimes n}$ has global symmetry $G^n \rtimes \S_n$, where $\S_n$ is the permutation symmetry, we can orbifold $\T^{\otimes n}$ by non-anomalous subgroups of $G^n \rtimes \S_n$.\footnote{
    In the absence of a global gravitational anomaly, the $\S_n$ permutation symmetry is non-anomalous \cite{Johnson-Freyd:2017ble}.
}
This way, we can verify the divisibility property \eqref{Divisibility} in a plethora of theories, supplying ample evidence for the claim of \cite{Lin:2021bcp}, and for the self-consistency of the Stolz--Teichner conjecture \cite{Segal,ST1,ST2}.
The relevant formulae for computing the torus partition functions of general orbifold theories are reviewed in {Section}~\ref{Sec:Orbifolds}.

The set of all $c=24$ bosonic holomorphic CFTs (with a unique vacuum) was conjectured by Schellekens \cite{Schellekens:1992db} (see also \cite{vanEkeren:2017scl}) and proven in \cite{Moller:2019tlx,Hohn:2020xfe,vanEkeren:2020rnz} except for the uniqueness of the Moonshine Module.
Their partition functions take the form $Z(\tau) = J(\tau) + 12\ell$, where $J$ is related to Klein's $j$-invariant by
\ie\label{J}
    J(\tau) = j(\tau) - 744 = \sum_{m\in\bZ} {d}(m) q^m = \frac{1}{q} + 196884 q + \cO(q^2),
\fe 
and
\ie\label{Schellekens}
    \ell \in 
    \{
        & 0,2,3,4,5,6,7,8,9,10,12,13,14,16,18,20,
        \\
        & 22,24,25,26,28,30,32,34,38,46,52,62,94\}.
\fe
Although some values of $\ell$ are realized by multiple theories, for the purpose of this note, there is no need to distinguish different theories with the same $\ell$, so we denote by $\Sch^\ell$ \emph{any} theory with $Z[\Sch^\ell](\tau) = J(\tau) + 12\ell$.

Our protagonist is the Monster CFT (Moonshine Module) $\M$ constructed by Frenkel, Lepowsky and Meurman \cite{frenkel1984natural,frenkel1988vertex}, which is a $c=24$ bosonic holomorphic CFT realizing the Monster group $\Mon$ as its global symmetry.
By focusing on $\M^{\otimes n}$ orbifolded by the \emph{full} $\S_n$ permutation symmetry mixed with subgroups of the diagonal $\Mon$, we find ourselves wielding the power of the \emph{second-quantized} formula of Dijkgraaf, Moore, Verlinde, Verlinde (DMVV) \cite{Dijkgraaf:1996xw}, together with the magic of the \emph{denominator} formula, originally due to Norton \cite{norton1984more} and later generalized by Borcherds \cite{borcherds1992monstrous}.  
The original denominator formula \cite{norton1984more} reads
\ie\label{Denom0}
    p^{-1} \prod_{n\in\bN,m\in\bZ} (1-p^n q^m)^{{d}(nm)} = J(\sigma) - J(\tau),
\fe 
where 
\ie 
    p := e^{2\pi i\sigma}
\fe 
throughout this note, and ${d}$ as defined in \eqref{J} are the Fourier coefficients of $J$.
Borcherds' \emph{twisted} denominator formula, given in \eqref{Denom}, replaces $J$ by the McKay--Thompson series.
We mainly consider orbifolds by \emph{cyclic} subgroups of $\Mon$ because the computation only requires the knowledge of the standard McKay--Thompson series, and not the generalized ones, some of which remain unknown.  One highlight of our exploration is a set of divisibility properties involving \emph{all the coefficients} of the McKay--Thompson series.  These results, together with a review of Monstrous Moonshine \cite{Conway:1979qga}, are presented in {Section}~\ref{Sec:TopMoonshine}.

Building on the unpublished work of Gaiotto and Yin \cite{GaiottoYin}, {Section}~\ref{Sec:Generalizations} studies generalizations to orbifolds by non-abelian subgroups of $\Mon$ and demonstrates the usage of the Generalized Moonshine \cite{norton1987generalized} data.  We then consider other theories in Schellekens' list \cite{Schellekens:1992db} and conclude with some remarks on future prospects.

In the remainder of this section, we introduce key tools for our exploration and present several number-theoretic results together with our main conjecture.

\subsection{Symmetric orbifolds and Kitaev chain}
\label{Sec:Arf}

Among all possible permutation orbifolds, the full $\S_n$ symmetric orbifold is special. 
The generating function
\ie\label{DMVV0}
    \cZ[\T](\sigma,\tau) &:= 1 + \sum_{n\in\bN} p^n Z[\T^{\otimes n}/\S_n](\tau)
\fe
admits a beautiful \emph{second-quantized} formula, which is a specialization of DMVV \cite{Dijkgraaf:1996xw}.
This second-quantized formula takes the form
\ie\label{DMVV1}
    \cZ[\T](\sigma,\tau) = \prod_{n\in\bN,m\in\bZ} (1-p^n q^m)^{-{d}(nm)},
\fe 
where 
\ie
    \cZ[\T](\tau) = \sum_{m\in\bZ} {d}(m) q^m.
\fe
In the case of the Monster CFT \cite{frenkel1984natural,frenkel1988vertex}, the famous denominator formula \cite{norton1984more}, whose left side is nothing but the inversion of the right side of \eqref{DMVV1}, gives a strikingly simple expression for $\cZ[\T](\sigma,\tau)$.
Although second-quantized formulae are also available for some other permutation orbifolds (e.g.\ symmetric orbifolds with discrete torsion \cite{Dijkgraaf:1999za}, and alternating orbifolds with or without discrete torsion \cite{Albert:2022gcs}), analogous denominator formulae do not exist.

We consider theories slightly more general than bosonic (non-spin) holomorphic CFTs, by allowing their tensor product with the fermionic (spin) invertible field theory $(-1)^\text{Arf}$, which arises as the low energy limit of the Kitaev chain \cite{Kitaev:2000nmw}.\footnote{
    An invertible field theory as introduced by Freed \cite{Freed:2014eja} is a topological quantum field theory with a unique vacuum.  The characterization of the Arf invariant as a mod 2 index was given by Atiyah \cite{atiyah1971riemann}.
}
By a slight abuse of language, we sometimes refer to the theory $(-1)^\text{Arf}$ simply as the Kitaev chain.
As a vertex operator superalgebra, the Neveu--Schwarz sector of $\tilde\T$ is isomorphic to $\T$ with purely bosonic (even) grading, and the Ramond sector (canonically twisted module) is isomorphic to $\T$ with purely fermionic (odd) grading.
For us, the relevant fact about $(-1)^\text{Arf}$ is that the torus partition functions of $\tilde\T := (-1)^\text{Arf} \otimes \T$ with even and odd spin structures are related to those of $\T$ by
\ie 
    Z_\text{even}[\tilde\T] = Z[\T], \quad Z_\text{odd}[\tilde\T] = Z_\text{ell}[\tilde\T] = -Z[\T],
\fe 
where $Z_\text{ell}$ denotes the elliptic genus.  
The divisibility property \eqref{Divisibility} with $Z(\tau)$ replaced by  $Z_\text{ell}(\tau)$ holds true for $c = 24n$ fermionic holomorphic CFTs, according to Stolz--Teichner \cite{Segal,ST1,ST2}.

Allowing tensor products with $(-1)^\text{Arf}$ serves two purposes:
\begin{enumerate}
    \item Since the divisibility property \eqref{Divisibility} holds true for theories with multiple vacua, we can realize partition functions $Z(\tau) = J(\tau) + 12\ell$ for any $\ell \in \bZ$ by considering direct sums of tensor products of Schellekens' theories \cite{Schellekens:1992db} with $(-1)^\text{Arf}$.
    For instance, $\ell = -1$ can be realized by $\Sch^2 \oplus \Sch^3 \oplus [ \Sch^6 \otimes (-1)^\text{Arf} ]$.
    Hence, by not demanding a unique vacuum, we could consider $c = 24$ partition functions with general $\ell \in \bZ$ in our orbifold constructions.
    \item It has been long noted that the DMVV formula \cite{Dijkgraaf:1996xw} applied to the Monster CFT \cite{frenkel1984natural,frenkel1988vertex} is related by an \emph{inversion} to the twisted denominator formula \cite{norton1984more,borcherds1992monstrous}.  Now, \emph{without inversion}, an alternative interpretation of the denominator formula is that it is the second-quantized elliptic genera of the tensor product of Monster with $(-1)^\text{Arf}$.\footnote{
        The twisted denominator formula \cite{norton1984more,borcherds1992monstrous} without inversion also figured in the work by Carnahan \cite{Carnahan:2009mjm}.
    }
\end{enumerate}

\subsection{Number theory and main conjecture}

We now present several number-theoretic results and the main conjecture.  Their relation to topological modular forms and the Moonshine Module $\M$ will be expatiated in {Section}~\ref{Sec:TopMoonshine}.

In discussing the divisibility property \eqref{Divisibility} for symmetric orbifolds, it is convenient to introduce the notion of \emph{series divisibility}.  Let $\bZ[[p]]^!$ denote the space of formal series in $p$ with unit constant coefficient.
\begin{definition}
    A series $S = 1 + \sum_{n=1}^\infty a_n p^n \in \bZ[[p]]^!$ is said to be divisible with base $D$ if
    \ie 
        \frac{D}{\mygcd{D,n}} \bdv a_n \text{~~for all~~} n \in \bN,
    \fe
    or equivalently,
    \ie 
        D \dv a_n n \text{~~for all~~} n \in \bN \quad\Leftrightarrow\quad D \bdv \frac{dS}{dp}.
    \fe
\end{definition}

\begin{lemma}
    \label{Closure}
    Series divisibility for a given base is preserved under multiplication and multiplicative inversion.
\end{lemma}
\begin{proof}
    Let $S, T \in \bZ[[p]]^!$ be divisible with base $D$, i.e.\
    \ie 
        D \bdv \frac{dS}{dp}, \frac{dT}{p}.
    \fe 
    Then the proposition simply follows from the Leibniz rule and the chain rule,
    \ie
        \frac{d(S T)}{dp} = \frac{dS}{dp} T + S \frac{dT}{dp},
        \quad
        \frac{d(S^{-1})}{dp} = - \frac{1}{S^2} \frac{dS}{dp}.
    \fe
\end{proof}

Note that in Lemma~\ref{Closure}, the preservation under multiplication holds more generally for $S, T \in \bZ[[p]]$ without assuming unit constant coefficient, however, that under multiplicative inversion requires $S \in \bZ[[p]]^!$ to ensure that $S^{-1} \in \bZ[[p]]^!$.

\begin{lemma}
    \label{Power}
    If $S \in \bZ[[p]]^!$ is divisible with base $D$, then $S^k$ is divisible with base $kD$ for $k \in \bZ$.
\end{lemma}
\begin{proof}
    This simply follows from the chain rule,
    \ie
        \frac{d(S^k)}{dp} = k S^{k-1} \frac{dS}{dp}.
    \fe
\end{proof}

Next we discuss the series divisibility of some elementary modular forms.  The Dedekind eta function of weight $\frac12$ is defined as
\ie 
    \eta = p^{\frac{1}{24}} \prod_{n=1}^\infty (1-p^n),
\fe 
which is related to the modular discriminant $\Delta$ (normalized to have unit leading coefficient) by $\Delta = \eta^{24}$.  Since $p^{-\frac{1}{24}} \eta \in \bZ[[p]]^!$ is tautologically divisible with base 1, we have the following corollary of Lemma~\ref{Power}.
\begin{corollary}
    \label{Eta}
    The series $p^{-\frac{k}{24}} \eta^k \in \bZ[[p]]^!$ for $k \in \bZ$ is divisible with base $k$.
\end{corollary}

The Eisenstein series is defined as
\ie 
    E_{2k} = 1 + \frac{2}{\zeta(1-2k)} \sum_{n=1}^\infty \frac{n^{2k-1} p^n}{1-p^n} \in \bZ[[p]]^!, \quad k \in \bN,
\fe
where $\zeta$ is the Riemann zeta function.  In particular, for $E_2$, $E_4$ and $E_6$,
\ie 
    \frac{2}{\zeta(-1)} = -24, \quad \frac{2}{\zeta(-3)} = 240, \quad \frac{2}{\zeta(-5)} = -504,
\fe 
which are all divisible by 24.  We arrive at the following lemma.
\begin{lemma}
    \label{Eisenstein}
    The Eisenstein series $E_{2k}$ for $k=1,2,3$ are divisible with base 24.  Moreover,
    \ie
        \frac{1-E_2}{24}, ~~ \frac{E_4-1}{240}, ~~ \frac{1-E_6}{504} \in p\bZ[[p]]^!.
    \fe
\end{lemma}

We are now ready to state a main proposition.  Klein's $j$-invariant can be defined as
\ie 
    j = \frac{E_4^3}{\eta^{24}} = \frac{1}{p} + 744 + 196884 p + \cO(p^2) \in \frac{\bZ[[p]]^!}{p},
\fe
and $J = j - 744$ is the trace function on the Moonshine Module $\M$ \cite{frenkel1984natural,frenkel1988vertex}.
\begin{proposition}
    \label{Main}
    The series
    \ie
        pj, ~~ pJ, ~~ -p^{-2} \left( \frac{dj}{dp} \right)^{-1} \in \bZ[[p]]^!
    \fe
    are divisible with base 24.
\end{proposition}
\begin{proof}
    The series divisibility of
    \ie
        pj = \frac{E_4^3}{p^{-1}\eta^{24}}
    \fe
    follows from Lemma~\ref{Closure}, Corollary~\ref{Eta} (for $k=24$) and Lemma~\ref{Eisenstein} (for $E_4$); that of $pJ$ is also true as it differs from $pj$ by a single term $744p = 24 \times 31 p$.  
    For the third series, because series divisibility is preserved under multiplicative inversion according to Lemma~\ref{Closure}, it suffices to show that 
    \ie\label{tmp}
        24 \bdv \frac{d}{dp} \left( p^2 \frac{dj}{dp} \right).
    \fe
    Under a trivial rewriting
    \ie 
        \frac{d}{dp} \left( p^2 \frac{dj}{dp} \right) 
        = p \frac{d^2(pj)}{dp^2},
    \fe
    \eqref{tmp} immediately follows from Proposition~\ref{Main} which says
    \ie 
        24 \bdv \frac{d(pj)}{dp}.
    \fe 
\end{proof}

Noting that $J$ is the McKay--Thompson series for the 1A class of the Monster group, Proposition~\ref{Main} admits a generalization involving the McKay--Thompson series for other conjugacy classes.  Let us set up the notation:
\begin{itemize}
    \item We follow the ATLAS \cite{abbott2015atlas} naming convention for conjugacy classes, in which classes of elements of order $n$ are named $nA, nB, nC, \dotsc$, in decreasing order of the size of the centralizer $\cen_G(g)$.
    \item The McKay--Thompson series of class $cl$ is denoted by $T_{cl} = p^{-1} + O(p) \in \bZ[[p]]^!/p$.
    \item The image of a class $cl$ under the $d$th power map is written as $cl^d$.
    \item Jordan's totient function $J_k(n)$ is defined as
\ie\label{Rho}
    J_k(n) = n^k \prod_{\text{prime } d \dv n} \left( 1 - \frac{1}{d^k} \right),
\fe
which is a count of the number of $k$-tuples of positive integers less than or equal to $n$, that together with $n$ form a coprime set of $k+1$ integers.
\end{itemize}

\begin{conjecture}
    \label{Conjecture}
    For every conjugacy class $cl$ of the Monster group, the series 
    \ie
        & p \left[ \sum_{d \dv N} \frac{J_2(N/d)}{N} T_{cl^d} \right] - (N-1),
        \\
        & \qquad\qquad\qquad\qquad - p^{-2} \left[ \sum_{d \dv N} \frac{J_2(N/d)}{N} \left( \frac{dT_{cl^d}}{dp} \right)^{-1} \right] - (N-1) \in \bZ[[p]]^!
    \fe
    are divisible with base 24, where $N$ is the order of elements in $cl$.
\end{conjecture}
The 1A case reduces to Proposition~\ref{Main}, while the other cases have been experimentally checked to high orders in $p$.

\section{Finite group and permutation orbifolds}
\label{Sec:Orbifolds}

Consider a CFT $\T$ in two spacetime dimensions.  The tensor product theory $\T^{\otimes n}$ has $\S_n$ permutation symmetry, and a \emph{permutation orbifold} is an orbifold of $\T^{\otimes n}$ by a subgroup $\Omega \le \S_n$.
For general $\Omega$, a formula for the torus partition function of $\T^{\otimes n}/\Omega$ was provided by Bantay \cite{Bantay:1997ek}.  For symmetric orbifolds $\Omega = \S_n$, a second-quantized formula for the generating function of $Z[\T^{\otimes n}/\S_n]$ was derived by Dijkgraaf, Moore, Verlinde, Verlinde (DMVV) \cite{Dijkgraaf:1996xw}.

If $\T$ has finite group global symmetry $G$, then $\T^{\otimes n}$ has global symmetry $G^n \rtimes \S_n$.  This section explains how to compute orbifolds by non-anomalous subgroups $H \times \Omega \le G_\text{diag} \times \S_n$, where $G_\text{diag}$ is the diagonal subgroup of $G^n$.
To this end, we will generalize Bantay's formula to allow twists by $H$, and review the equivariant second-quantized formula for symmetric orbifolds due to Tuite \cite{Tuite:2008gy}.

\subsection{Partition functions under orbifolds}

Let $G$ be a non-anomalous symmetry group, and $Z_h^g$ denote the torus partition functions twisted by $g$ in time and $h$ in space, where $g$ and $h$ commute.\footnote{
    In terms of topological defect lines, the $h$ defect runs upwards in the time direction, the $g$ defect runs toward the right in the space direction, and the group multiplication at a vertex reads
    \[
        \begin{gathered}
            \begin{tikzpicture}[scale=0.5]
            \draw [line,->-=0.6] (0,0) -- (0,1) node [above] {$gh$};
            \draw [line,-<-=0.6] (0,0) -- (-.87,-.5) node [below left] {$g$};
            \draw [line,-<-=0.6] (0,0) -- (.87,-.5) node [below right] {$h$};
            \end{tikzpicture}
        \end{gathered}.
    \]
}
We adopt the convention that under modular transformations,\footnote{
    While many conventions exist in the literature, e.g.\ Dijkgraaf, Vafa, Verlinde, Verlinde \cite[(4.18)]{Dijkgraaf:1989hb}, or Dong and Mason \cite[(1.4)]{Dong:1994wn}, we choose to follow Tuite \cite[(14)]{Tuite:2008gy} closely since his paper contains many useful formulae for our computations.
}
\ie\label{ModularConvention}
    \begin{pmatrix}
        a & b \\ c & d
    \end{pmatrix} \in {\rm SL}(2,\bZ):~~Z_h^g(\tau) = Z^{g^a h^b}_{g^c h^d}(\frac{a\tau+b}{c\tau+d}).
\fe
Another useful identity is $Z_h^g = Z_{xhx^{-1}}^{xgx^{-1}}$ for any pair of commuting $g, h \in G$ and any $x \in G$.\footnote{
    In terms of topological defect lines, we nucleate a loop of $x \in G$ on the torus, and perform the obvious moves to turn $Z_h^g$ into $Z_{xhx^{-1}}^{xgx^{-1}}$.
}
It follows that $Z^g$ and $Z_g$ only depend on the conjugacy class of $g$.

Orbifolding \cite{Dixon:1985jw,Dixon:1986jc} by $G$ refers to the procedure of adding twisted sectors, one for each element $h \in G$, and projecting to the singlet of each centralizer $\cen_G(h)$ by averaging over $g \in \cen_G(h)$.  The result is \cite{Dijkgraaf:1989hb}
\ie\label{ZOrbifold}
    Z[\T/G] &= \frac{1}{|G|} \sum_{h \in G} \sum_{g  \in \cen_G(h)} Z[\T]_h^g = \sum_{cl = [h] \in \cl_G} \frac{1}{|\cen_G(h)|} \sum_{g \in \cen_G(h)} Z[\T]_h^g,
\fe 
where $\cl_G$ denotes the set of all conjugacy classes of $G$, $cl$ is a conjugacy class, which is alternatively written as $[h]$ if $h$ is a representative of the class, and the basic identity $|G| = |[h]| |\cen_G(h)|$ was used.
We can also treat space and time on more equal footing to manifest modular invariance by expressing the above as
\ie\label{T/G}
    Z[\T/G] = \frac{1}{|G|} \sum_{\substack{g,h \\ \comm{g}{h}}} Z[\T]_h^g.
\fe

\subsection{Cyclic group orbifolds}
\label{Sec:Cyclic}

When $G = \bZ_N = \{0, 1, \dotsc, N-1\}$, the general orbifold formula \eqref{T/G} becomes\footnote{
    Since group elements are labeled by integers in this subsection, to avoid confusion with the $r$th power of $Z$, we write $Z^r_0$ instead of $Z^r$ to denote the partition function with a temporal twist and no spatial twist.
}
\ie\label{Cyclic}
    Z[\T/\bZ_N] = \frac{1}{N} \sum_{r,s=0}^{N-1} Z[\T]^r_s.
\fe
Let $p := r / \mygcd{r,s,N}$ and $q := s / \mygcd{r,s,N}$. 
It follows that $\mygcd{p,q,N} = 1$, which
implies the existence of an integer $t$ such that $\mygcd{p+tN,q} = 1$.
By setting $g = \mygcd{r,s,N}$, $h = 0$, $a = p+tN$, $c = q$ (and $b,d$ such that $ad-bc=1$) in \eqref{ModularConvention}, we see that $Z[\T]^r_s$ is in the modular orbit of $Z[\T]^{\mygcd{r,s,N}}_0$.
Because modular transformations leave $\mygcd{r,s,N}$ invariant, we conclude that there are precisely $\sigma_0(N)$---the number of divisors of $N$---modular orbits.
For each divisor $d \dv N$, a representative can be chosen to be $Z[\T]^d_0$, and there are $J_2(N/d)$ elements in the modular orbit, where $J_k(n)$ is Jordan's totient function defined in \eqref{Rho}.
In particular, if $N$ is prime, then
\ie\label{RhoPrime}
    J_2(N/d) = \begin{cases}
        1, & d=N,
        \\
        N^2-1, & d=1.
    \end{cases}
\fe
Sometimes, $\T$ and $\bZ_N$ exhibit the property that
\ie\label{PowerMapProperty}
    Z[\T]^r_0 = Z[\T]^{\mygcd{r,N}}_0 \text{~~for all~~} r,
\fe
in which case several partition functions in each modular orbit are identical, and the orbifold computations get simplified.\footnote{
    A sufficient condition for this property is if the outer automorphism group $\text{Aut}(\bZ_N$) is a symmetry.  Note that the CPT symmetry implies $Z[\T]^r_0 = Z[\T]^{N-r}_0$.  The author thanks Yifan Wang for a discussion.
}

When $N$ is prime, we can write \eqref{Cyclic} as
\ie\label{CyclicPrime}
    Z[\T/\bZ_N](\tau) &= \frac{1}{N} \left( \sum_{r=0}^{N-1} Z[\T]^r_0(\tau) + \sum_{s=1}^{N-1} \sum_{r=0}^{N-1} Z[\T]_s(\tau+r) \right)
    \\
    &= \frac{1}{N} \sum_{r=0}^{N-1} Z[\T]^r_0(\tau) + \sum_{s=1}^{N-1} Z[\T]_s(\tau)\big|_{q^\bZ},
\fe 
where $|_{q^\bZ}$ means projection to integral Fourier powers. 
If $\T$ and $\bZ_N$ further exhibit property \eqref{PowerMapProperty}, then we can reduce \eqref{CyclicPrime} to
\ie\label{CyclicPrimeSpecial}
    Z[\T/\bZ_N](\tau) = \frac{1}{N} Z[\T](\tau) + \frac{N-1}{N} Z[\T]^1_0(\tau) + (N-1) Z[\T]_1(\tau)\big|_{q^\bZ}.
\fe

\subsection{Permutation orbifolds}

A general formula for the orbifold partition function $Z[\T^{\otimes n}/\Omega]$ by any permutation group $\Omega \le \S_n$ was provided by Bantay \cite{Bantay:1997ek}.  With small modifications, the twisted partition functions $Z[\T^{\otimes n}/\Omega]_h^g$ by elements $g,h$ of a non-anomalous subgroup $H < G_\text{diag}$ can also be computed.  In order to present the formula, we need to first define some quantities by considering $\Omega$ acting on the point set $\{1, 2, \dotsc, n\}$.
Suppose $x, y \in \Omega$ and $\comm{x}{y}$, then $x$ and $y$ generate a subgroup $\langle x, y \rangle \le \Omega$. 
Under $\langle x, y \rangle$, the point set $\{1, 2, \dotsc, n\}$ organizes into orbits, the collection of which we denote by $O(x,y)$.  For each orbit $\xi \in O(x,y)$, the elements can be further decomposed into orbits of $\langle x \rangle$.  Define the following quantities:
\begin{enumerate}
    \item  $\mu_\xi$: the number of $\langle x \rangle$ orbits in a given $\langle x,y \rangle$ orbit $\xi$.
    \item  $\lambda_\xi$: the length of any $\langle x \rangle$ orbit in a given $\langle x,y \rangle$ orbit $\xi$.  Note the relation $\mu_\xi \lambda_\xi = |\xi|$.
    \item  $\kappa_\xi$: the smallest non-negative integer for which $y^{\mu_\xi} = x^{\kappa_\xi}$.
\end{enumerate}
Then, the twisted partition functions are given by
\ie\label{Bantay}
    Z[\T^{\otimes n}/\Omega]_h^g(\tau) = \frac{1}{|\Omega|} \sum_{\substack{x,y\in\Omega \\ \comm{x}{y}}} \prod_{\xi \in O(x,y)} Z[\T]^{g^{\mu_\xi} h^{\kappa_\xi}}_{h^{\lambda_\xi}}(\tau_\xi), \quad \tau_\xi = \frac{\mu_\xi \tau + \kappa_\xi}{\lambda_\xi}.
\fe

\subsection{Symmetric orbifolds}
\label{Sec:SecondQuantized}

Further magic happens when we consider the full $\S_n$ symmetric orbifold.
Given a non-anomalous subgroup $H < G_\text{diag}$ and a pair $g, h \in H$, it is natural to define a generating function for the symmetric orbifold partition functions,
\ie
    \cZ[\T]_h^g(\sigma,\tau)
    := 1 + \sum_{n\in\bN} p^n Z[\T^{\otimes n}/\S_n]^g_h(\tau).
\fe
Tuite \cite{Tuite:2008gy}, generalizing Bantay \cite{bantay2003symmetric}, provided an expression for the above
in terms of twisted (equivariant) Hecke operators \cite{baker1990hecke,borcherds1992monstrous,Tuite:2008gy,ganter2009hecke,carnahan2010generalized}, as follows.
The $n$th equivariant (twisted) Hecke operator $T_n$ acting on a weight-zero modular function is defined as \cite[(15)]{Tuite:2008gy}
\ie 
    T_n Z_h^g(\tau) = \frac{1}{n} \sum_{\substack{ad=n \\ 0 \le b < d}} Z^{g^a h^b}_{h^d} \left( \frac{a\tau + b}{d} \right).
\fe
Then, \cite[(35)]{Tuite:2008gy}
\ie\label{Hecke}
    \cZ[\T]_h^g(\sigma,\tau)
    = \exp\left\{ \sum_{n\in\bN} p^n T_n Z[\T]_h^g(\tau) \right\}.
\fe

Formula \eqref{Hecke} with no spatial twist, $h=e$, reduces to the following second-quantized formula of Dijkgraaf, Moore, Verlinde, Verlinde (DMVV) \cite{Dijkgraaf:1996xw}.
Suppose $g$ is of order $N$, and let $d^g(m,\ell)$ denote the number of states in the single-copy theory $\T$ with $(L_0 - \frac{c}{24})$-eigenvalue $m$ and $g$-eigenvalue $e^{\frac{2\pi i \ell}{N}}$,
such that
\ie\label{d}
    Z[\T]^{g^k}(\tau) &= \sum_{m\in\bZ,\ell\in\bZ_{N}} d^g(m,\ell) q^m e^{\frac{2\pi i k \ell}{N}}, \quad k \in \bZ_{N}.
\fe
Then
\ie\label{DMVV}
    \cZ[\T]^g(\sigma,\tau) &= \prod_{n\in\bN,m\in\bZ,\ell\in\bZ_{N}} \frac{1}{(1-p^n q^m e^{\frac{2\pi i \ell}{N}})^{d^g(nm,\ell)}}.
\fe
The next section connects \eqref{DMVV} to the twisted denominator formula \cite{norton1984more,borcherds1992monstrous} when $\T$ is the Monster CFT \cite{frenkel1984natural,frenkel1988vertex}.

\section{Topological modularity of Monstrous Moonshine}
\label{Sec:TopMoonshine}

\subsection{Monstrous Moonshine}
\label{Sec:Moonshine}

The Monster CFT (Moonshine Module) $\M$ is a $c=24$ bosonic holomorphic CFT constructed by Frenkel, Lepowsky and Meurman \cite{frenkel1984natural,frenkel1988vertex}, which realizes the Monster group $\Mon$ as its global symmetry.
The existence of $\M$ explains the observation that the Fourier coefficients of $Z[\M] = J$ are simple summations of the dimensions of irreducible representations of $\Mon$.
The twisted partition functions $Z[\M]^g$ are called the McKay--Thompson series, whose functional forms were first conjecturally enumerated by Conway and Norton \cite{Conway:1979qga}, and later proven by Borcherds \cite{borcherds1992monstrous}; their Fourier coefficients are also simple summations of the characters of $\Mon$.
There are 171 distinct McKay--Thompson series for 194 conjugacy classes, since certain pairs of distinct conjugacy classes, e.g.\ 23A and 23B, share identical McKay--Thompson series.\footnote{
    Conjugacy classes related by inversion must share the same McKay--Thompson series, as decreed by the CPT symmetry.  All coincidences of the McKay--Thompson series can be explained this way, except for 27AB.  The author thanks Yifan Wang for a discussion.
}
Monstrous Moonshine \cite{Conway:1979qga} refers to a collection of fascinating properties of $\M$ and the McKay--Thompson series.\footnote{
    See \cite{Paquette:2016xoo,Paquette:2017xui} for a physics derivation of Monstrous Moonshine.  The most nontrivial aspect of the Monstrous Moonshine \cite{Conway:1979qga} is that the McKay--Thompson series are Hauptmoduls of genus-zero subgroups of $\text{PSL}(2,\bR)$.  Since this property plays little role in the present pursuit, we make no further mention.
}

Johnson-Freyd \cite{Johnson-Freyd:2017ble} showed that the Monster CFT $\M$ has an anomaly of exact order 24 in $\H^3(\Mon,\text{U}(1))$.\footnote{
    Johnson-Freyd and Treumann \cite{johnson2019third} further showed that the $\bZ_{24}$ generated by $\M$ is a direct summand of $\H^3(\Mon,\text{U}(1))$, and $\H^3(\Mon,\text{U}(1)) \ominus \bZ_{24}$ has order dividing 4.  The author thanks Theo Johnson-Freyd for explaining these results.
}
This anomaly manifests itself in the modular properties of the McKay--Thompson series.  The anomaly of any subgroup of $\Mon$ is given by the pullback, and consistent orbifold operations can be performed only for non-anomalous subgroups.  In particular, for the cyclic subgroup $\vev{g}$ generated by $g \in \Mon$ of order $N$, if $Z[\M]^g(\frac{\tau}{1-N\tau}) = e^{\frac{2\pi i k}{N}} Z[\M]^g(\tau)$, then $\vev{g}$ has $k \pmod N$ units of the $\bZ_N$ anomaly \cite{Lin:2021udi}; clearly, the anomaly only depends on the conjugacy class of $g$.

\begin{table}[t]
    \centering
    \begin{tabular}{|>{\centering\arraybackslash}p{3cm}|>{\centering\arraybackslash}p{8cm}|>{\centering\arraybackslash}p{4cm}|}
        \hline
        & Fricke & Non-Fricke
        \\\hline
        \qquad\qquad\qquad\qquad \parbox{2cm}{} \qquad\qquad\qquad\qquad \parbox{2cm}{}
        Anomalous
        & 
        3C, 4B, 8B, 8C, 12C, 12D, 12F, 16A, 20B, 20E, 21C, 24A, 24E, 24F, 24H, 28A, 28D, 30E, 32B, 36C, 39B, 40A, 40B, 40CD, 48A, 52A, 52B, 56BC, 57A, 60A, 60E, 60F, 68A, 84A, 84C, 88AB, 93AB, 104AB 
        & 
        4D, 6F, 8D, 8F, 12G, 12J, 15D, 20D, 24D, 24G, 24J, 42C, 84B
        \\\hline
        \qquad\qquad\qquad\qquad \parbox{2cm}{} \qquad\qquad\qquad\qquad \parbox{2cm}{}
        \qquad\qquad\qquad\qquad \parbox{2cm}{} \qquad\qquad\qquad\qquad \parbox{2cm}{}
        \qquad\qquad\qquad\qquad \parbox{2cm}{}
        Non-anomalous 
        &
        1A, 2A, 3A, 4A, 5A, 6A, 6B, 7A, 8A, 9A, 10A, 10D, 11A, 12A, 12H, 13A, 14A, 14C, 15A, 15C, 16C, 17A, 18B, 18E, 19A, 20A, 20F, 21A, 21D, 22A, 23AB, 24B, 24I, 25A, 26A, 26B, 27AB, 28B, 29A, 30B, 30D, 30F, 31AB, 32A, 33B, 34A, 35A, 35B, 36A, 36D, 38A, 39A, 39CD, 41A, 42A, 42D, 44AB, 45A, 46CD, 47AB, 50A, 51A, 54A, 55A, 56A, 59AB, 60B, 60C, 62AB, 66A, 66B, 69AB, 70A, 71AB, 78A, 87AB, 92AB, 94AB, 95AB, 105A, 110A, 119AB
        &
        2B, 3B, 4C, 5B, 6C, 6D, 6E, 7B, 8E, 9B, 10B, 10C, 10E, 12B, 12E, 12I, 13B, 14B, 15B, 16B, 18A, 18C, 18D, 20C, 21B, 22B, 24C, 28C, 30A, 30C, 30G, 33A, 36B, 42B, 46AB, 60D, 70B, 78BC
        \\\hline
    \end{tabular}
    \\
    ~\vspace{.3in}~
    \\
    \begin{tabular}{|c|c|c|c|}
        \hline
        & Fricke & Non-Fricke & Total
        \\\hline
        Anomalous & 43/38 & 13/13 & 56/51
        \\\hline
        Non-anomalous & 98/82 & 40/38 & 138/120
        \\\hline 
        Total & 141/120 & 53/51 & 194/171
        \\\hline
    \end{tabular}
    \\
    ~\vspace{.1in}~
    \\
    \caption{Organizing the 194 conjugacy classes/171 distinct McKay--Thompson series of the Monster Group in the Monster CFT according to whether $g$ is anomalous/Fricke.}
    \label{Tab:Fricke}
\end{table}

The McKay--Thompson series for certain classes are invariant under the so-called \emph{Fricke involutions},
\ie 
    Z[\M]^g(\tau) \mapsto Z[\M]^g(-\frac{1}{\A N\tau}),
\fe
where $N$ is the order of $g$ and $\A = N/\mygcd{N,k}$, giving
\ie\label{Fricke}
    Z[\M]_g(\tau) = Z[\M]^g(-\frac{1}{\tau}) = Z[\M]^g(\frac{\tau}{\A N}),
\fe
where the first equality is due to modular covariance \eqref{ModularConvention}.
An element $g \in \Mon$ satisfying \eqref{Fricke} is said to be \emph{Fricke}.
It is known that $g$ is Fricke if and only if $Z[\M]^g(\tau)$ has a pole at $\tau = 0$, or equivalently, $Z[\M]_g(\tau)$ has a pole at $\tau = i\infty$ \cite{borcherds1992monstrous}.
Table~\ref{Tab:Fricke} organizes the 194 conjugacy classes/171 distinct McKay--Thompson series according to whether $g$ is anomalous/Fricke.

Finally, we note that every Monster group element satisfies property \eqref{PowerMapProperty}.\footnote{
    While it is not true that the conjugacy classes $\cl(g^r)$ only depend on $\mygcd{N,r}$ for every Monster group element $g$, it is true up to inversion.  For instance, $g\in\text{23A}$ and $g^2\in\text{23B}$, despite $\mygcd{23,1} = \mygcd{23,2} = 1$; however, because $g^{-1}\in\text{23B}$, the conjugacy classes 23AB are related by inversion, and must share the same McKay--Thompson series by the CPT symmetry.  The author thanks Yifan Wang for a discussion.
}

\subsection{Denominator formula}
\label{Sec:Denominator}

The DMVV formula \cite{Dijkgraaf:1996xw} \eqref{DMVV} for symmetric orbifolds of the Monster CFT $\M$ has a strikingly simple expression given by the inverse of the twisted denominator formula \cite{norton1984more,borcherds1992monstrous}.
For any $g \in \Mon$ of order $N$,
\ie\label{Denom}
    p^{-1} \cZ[\M]^g(\sigma,\tau)^{-1} = p^{-1} \prod_{n\in\bN,m\in\bZ,\ell\in\bZ_{N}} (1-p^n q^m e^{\frac{2\pi i \ell}{N}})^{d^g(nm,\ell)} = Z[\M]^g(\sigma) - Z[\M]^g(\tau),
\fe 
where $d^g(\ell,m)$ was defined in \eqref{d}.
There are two physical interpretations of this formula:
\begin{enumerate}
    \item Regarding $-Z[\M]^g = Z_\text{ell}[\tilde\M = \M \otimes (-1)^{\text{Arf}}]^g$ as the time-twisted elliptic genera of the Monster CFT coupled to the Kitaev chain \cite{Kitaev:2000nmw}, we interpret formula \eqref{Denom} \emph{without inversion} as encoding the time-twisted elliptic genera of the symmetric orbifolds of $\tilde\M$.
    \item Inverting \eqref{Denom}, $(Z[\M]^g(\sigma) - Z[\M]^g(\tau))^{-1}$ becomes a generating function for the twisted partition functions of the symmetric orbifolds of $\M$.\footnote{
        It is important to series expand first in $p$ and then in $q$.
    }
\end{enumerate}
The beauty of the twisted denominator formula is that now concise formula can be written down for the elliptic genera (resp.\ partition functions) of orbifolds of $\tilde\M^{\otimes n}/\S_n$ (resp.\ $\M^{\otimes n}/\S_n$) by non-anomalous (resp.\ non-anomalous, Fricke, and prime) cyclic subgroups of $\Mon$.

\subsection{Symmetric orbifolds of the Monster coupled to Kitaev chain}

We begin with the symmetric orbifolds of the Monster CFT coupled to the Kitaev chain \cite{Kitaev:2000nmw}, $\tilde\M = \M \otimes (-1)^{\text{Arf}}$.  Suppose $g \in \Mon$ is a non-anomalous element of order $N$.
By virtue of the twisted denominator formula \eqref{Denom}, the $g$-twisted genera of symmetric orbifolds for $n>1$ copies are simply \emph{constants} and given by the $p^{n-1}$ coefficients of $Z[\M]^g(\sigma)$,
\ie
    Z_\text{ell}[\tilde\M^{\otimes n}/\S_n]^g = Z[\M]^g(\sigma)\big|_{p^{n-1}}.
\fe
From the discussions about modular orbits in {Section}~\ref{Sec:Cyclic},
we deduce that the orbifold theories $\tilde\M^{\otimes n}/\S_n \times \vev{g}$ with $n>1$ have elliptic genera
\ie 
    Z_\text{ell}[\tilde\M^{\otimes n}/\S_n \times \vev{g}] &= \sum_{d \dv N} \frac{J_2(N/d)}{N} Z_\text{ell}[\tilde\M^{\otimes n}/\S_n]^{g^d}
    \\
    &
    = 
    \left[ \sum_{d \dv N} \frac{J_2(N/d)}{N} Z[\M]^{g^d}(\sigma) \right]_{p^{n-1}},
\fe 
where $J_2(N/d)$ is Jordan's totient function defined in \eqref{Rho}.
These elliptic genera can be further encapsulated by a generating function
\ie\label{GeneratingArf}
    \cZ_\text{ell}[\tilde\M;g](\sigma) &:= 1 + \sum_{n>1} p^n Z_\text{ell}[\tilde\M^{\otimes n}/\S_n \times \vev{g}] 
    \\
    &= p \left[ \sum_{d \dv N} \frac{J_2(N/d)}{N} Z[\M]^{g^d}(\sigma) \right] - (N-1).
\fe 
The divisibility property \eqref{Divisibility} states that $\cZ_\text{ell}[\tilde\M;g](\sigma)$ should be divisible with base 12.  However, we empirically observe that $\cZ_\text{ell}[\tilde\M;g](\sigma)$ is divisible with base 24.

\subsection{Symmetric orbifolds of the Monster}
\label{Sec:SymMon}

Next, we consider symmetric orbifolds of the Monster CFT $\M$.  Suppose $g \in \Mon$ is a non-anomalous element of order $N$, and let
\ie 
    Z[\M^{\otimes n}/\S_n]^g(\tau) = \sum_{m=-n}^\infty a^g(n,m) q^m.
\fe
The coefficients $a(n,m)$ can be extracted by 
\ie
     a^g(n,m) &= \frac{1}{(2\pi i)^2} \oint \frac{dq}{q^{m+1}} \oint \frac{dp}{p^{n+2}} \frac{1}{Z[\M]^g(\sigma)-Z[\M]^g(\tau)}
     \\
     &= \frac{1}{(2\pi i)^2} \oint \frac{dq}{q^{m+1}} \oint \frac{dp}{p^{n+2}} \frac{1}{p-q} \frac{p-q}{Z[\M]^g(\sigma)-Z[\M]^g(\tau)}
     \\
     &= - \frac{1}{2\pi i} \oint \frac{dq}{q^{n+m+3}} \left( \frac{dZ[\M]^g(\tau)}{dq} \right)^{-1}
     \\
     & \qquad + \frac{1}{(2\pi i)^2} \oint \frac{dp}{p^{n+2}} \oint \frac{dq}{q^{m+1}} \frac{1}{Z[\M]^g(\sigma)-Z[\M]^g(\tau)},
\fe
where in the last step we commuted the two integrals and picked up the residue at $p=q$.
Note that the second term in the final expression vanishes for $m\le0$.  Hence, for $m\le0$, $a^g(n,m)$ only depends on $n+m$, and we can write $a^g(n,m) = b^g(n+m)$, where
\ie 
    \left( \frac{dZ[\M]^g(\tau)}{dq} \right)^{-1} = - \sum_{n=0}^\infty b^g(n) q^{n+2}.
\fe
In light of the divisibility property \eqref{Divisibility}, we are particularly interested in the constant Fourier coefficient, 
\ie 
    Z[\M^{\otimes n}/\S_n]^g(\tau)\big|_{q^0} = a^g(n,0) = b^g(n) = - \left( \frac{dZ[\M]^g(\tau)}{dq} \right)^{-1} \bigg|_{q^{n+2}}.
\fe 

If $g$ is Fricke in $\M$, i.e.\ \eqref{Fricke} is satisfied, then $g$ is Fricke in all symmetric orbifolds of $\M$, in the sense that \eqref{Fricke} is true with $\M$ replaced by $\M^{\otimes n}/\S_n$.  This can be seen by invoking the twisted denominator formula \eqref{Denom} to obtain
\ie
    \cZ[\M]_g(\sigma,\tau) = \cZ[\M]^g(\sigma,-\frac{1}{\tau}) = \cZ[\M]^g(\sigma,\frac{\tau}{\A N}).
\fe
It then follows that
$Z[\M^{\otimes n}/\S_n]^g(\tau)$ and $Z[\M^{\otimes n}/\S_n]_g(\tau) = Z[\M^{\otimes n}/\S_n]^g(-1/\tau)$ share identical constant Fourier coefficients,
\ie\label{ConstFricke}
    Z[\M^{\otimes n}/\S_n]^g(\tau)\big|_{q^0} = Z[\M^{\otimes n}/\S_n]_g(\tau)\big|_{q^0} \text{~~for all~~} n.
\fe

Now, suppose $g$ is a non-anomalous Fricke element of \emph{prime} order $N$.  Combining \eqref{CyclicPrime} and \eqref{ConstFricke} gives a simple formula for the constant Fourier coefficients under orbifolds by $\vev{g}$,
\ie 
    Z[\M^{\otimes n}/\S_n \times \vev{g}](\tau)\big|_{q^0} = \frac{1}{N} Z[\T](\tau)\big|_{q^0} + \frac{N^2-1}{N} Z[\T]^g(\tau)\big|_{q^0}.
\fe 
These can be encapsulated by a generating function
\ie\label{GeneratingMonsterPrimeFricke}
    \cZ^\text{const}[\M;g](\sigma) &:= 1 + \sum_{n>1} p^n Z[\M^{\otimes n}/\S_n \times \vev{g}]_{q^0}
    \\
    & \hspace{-.5in} = - p^{-2} \left[ \frac{1}{N} \left( \frac{dZ[\M](\sigma)}{dp} \right)^{-1} + \frac{N^2-1}{N} \left( \frac{dZ[\M]^g(\sigma)}{dp} \right)^{-1} \right] - (N-1).
\fe 
The divisibility property \eqref{Divisibility} states that 
$\cZ^\text{const}[\M;g](\sigma)$ should be divisible with base 12.  However, we empirically observe that $\cZ^\text{const}[\M;g](\sigma)$ is divisible with base 24.

\subsection{Series divisibility of the McKay--Thompson Series}

Explorations in the previous two subsections resulted in statements of series divisibility involving a \emph{subset} of the McKay--Thompson series.  We now extend the statements to \emph{all} the McKay--Thompson series.

Why were they a subset?  Because orbifolds only make physical sense for non-anomalous symmetry groups.
If $g$ has $k \pmod N$ units of the $\bZ_N$ anomaly in the single-copy Monster CFT $\M$, then the $\bZ_N$ orbifold of $n$ copies is only consistent if the diagonal $\bZ_N$ is non-anomalous, i.e.\ $N \dv nk$.  Hence, on physical grounds, the divisibility property \eqref{Divisibility} only needs to be satisfied for $n$ a multiple of $N/\mygcd{N,k}$.  For instance, the 3C class is anomalous, so the 3C orbifold is consistent only when $3 \dv n$.
Nevertheless, both $\cZ_\text{ell}[\tilde\M;\text{3C}]$ and $\cZ^\text{const}[\M;\text{3C}]$ defined in \eqref{GeneratingArf} and \eqref{GeneratingMonsterPrimeFricke} turn out to be divisible with base 24.  

Let us explicitly demonstrate the series divisibility of the 1A, 2A, and 3C McKay--Thompson series,
\ie 
    \cZ_\text{ell}[\tilde\M;\text{1A}](\sigma) &= p J(\sigma)
    \\
    &= 1+196884 p^2+21493760 p^3+864299970 p^4+20245856256 p^5
    \\
    &\qquad +333202640600 p^6+4252023300096 p^7+44656994071935 p^8
    \\
    &\qquad +401490886656000 p^9+3176440229784420
    p^{10}
    \\
    &\qquad +22567393309593600 p^{11}+146211911499519294 p^{12} + \cO(p^{13}),
    \\
    \cZ_\text{ell}[\tilde\M;\text{2A}](\sigma) &= p 
    \left( \frac12 J(\sigma) + \frac32 Z[\M]^\text{2A}(\sigma) \right) - 1
    \\
    &= 1+105000 p^2+10891264 p^3+434009988 p^4+10138976256 p^5
    \\
    &\qquad +166712962480 p^6+2126658945024 p^7+22331807148798 p^8
    \\
    &\qquad +200760619696128 p^9+1588284040335048
    p^{10}
    \\
    &\qquad +11283946500956160 p^{11}+73106873450072700 p^{12} + \cO(p^{13}),
    \\
    \cZ_\text{ell}[\tilde\M;\text{3C}](\sigma) &= p 
    \left( \frac13 J(\sigma) + \frac83 Z[\M]^\text{3C}(\sigma) \right) - 2
    \\
    &= 1+65628 p^2+7165248 p^3+288099990 p^4+6748618752 p^5
    \\
    &\qquad +111067557864 p^6+1417341100032 p^7+14885664690645 p^8
    \\
    &\qquad +133830295644672 p^9+1058813409928140 p^{10}
    \\
    &\qquad +7522464436531200 p^{11}+48737303833741434 p^{12} + \cO(p^{13}).
\fe 
If we divide the $p^n$ coefficient by $12/\mygcd{12,n}$ for $n \ge 1$, then we get
\ie 
    1A: \quad & 0,32814,5373440,288099990,1687154688,166601320300,354335275008,
    \\
    & 14885664690645,100372721664000,529406704964070,1880616109132800,
    \\
    & 146211911499519294, \dotsc,
    \\
    2A: \quad & 0,17500,2722816,144669996,844914688,83356481240,177221578752,
    \\
    & 7443935716266,50190154924032,264714006722508,940328875079680,
    \\
    & 73106873450072700, \dotsc,
    \\
    3C: \quad & 0,10938,1791312,96033330,562384896,55533778932,118111758336,
    \\
    & 4961888230215,33457573911168,176468901654690,626872036377600,
    \\
    & 48737303833741434, \dotsc,
\fe 
which are all integers.  For the non-anomalous 2A, this is the requirement of the divisibility property \eqref{Divisibility} applied to $\tilde T^{\otimes n}/\S_n \times \vev{\text{2A}}$.
For the anomalous 3C, the divisibility of every third Fourier coefficient is demanded by \eqref{Divisibility}, but the divisibility of the full series is not.

In the case of the Monster CFT $\M$ without $(-1)^\text{Arf}$, the generating function \eqref{GeneratingMonsterPrimeFricke} is valid only if $g$ is a non-anomalous Fricke element of prime order.  For instance, consider the non-anomalous but non-Fricke 2B class.  A direct orbifold computation shows that for $n \ge 1$,
\ie 
    Z[\M^{\otimes n}/\S_n \times \vev{\text{2B}}]_{q^0} &= 24, 98880, 21494336, 20678177700, 8504045476896,
    \\
    & \qquad 5251234688999200, \dotsc,
\fe
whereas \eqref{GeneratingMonsterPrimeFricke} gives\footnote{
    For a single copy $n=1$, it is well-known that $\M/\vev{p\text{B}}$ with $p=2,3,5,7,13$ gives the Leech lattice CFT, with 24 spin-one conserved currents coming from the twisted sector; by contrast, \eqref{GeneratingMonsterPrimeFricke} gives none.
}
\ie 
    0,98856,21487616,20678269356,8504042913792,5251234745733328,\dotsc.
\fe
It turns out that the differences, albeit nonzero, are divisible by $12/\mygcd{12,n}$, with quotients
\ie 
    2, 4, 1680, -30552, 213592, -28367064, \dotsc.
\fe

Given that divisibility seems to hold beyond the scope decreed by physics, we consider a further generalization.
Define for any $g \in \Mon$,
\ie\label{GeneratingMonster}
    \cZ^\text{const}[\M;g](\sigma) &:= 1 + \sum_{n>1} p^n Z[\M^{\otimes n}/\S_n \times \vev{g}]_{q^0}
    \\
    &= - p^{-2} \left[ \sum_{d \dv N} \frac{J_2(N/d)}{N} \left( \frac{dZ[\M]^{g^d}(\sigma)}{dp} \right)^{-1} \right] - (N-1),
\fe 
which reduces to the physical generating function \eqref{GeneratingMonsterPrimeFricke} when $g$ is a non-anomalous Fricke element of prime order.  Clearly, this definition borrows the form of \eqref{GeneratingArf}.

We empirically observe that both $\cZ_\text{ell}[\tilde\M;g]$ and $\cZ^\text{const}[\M;g]$ defined in \eqref{GeneratingArf} and \eqref{GeneratingMonster} are divisible with base \emph{24} (not just 12) for \emph{every} $g \in \Mon$.
This observation is the basis of our main Conjecture~\ref{Conjecture}; in particular, Proposition~\ref{Main} concerns the proven special case of $g=e$.
We emphasize again that Conjecture~\ref{Conjecture} is strictly stronger than what is decreed by physical consistency---not only is the base doubled from 12 to 24 but the non-anomalous/Fricke conditions are also relaxed.
The divisibility of the McKay--Thompson series provides a nontrivial check of \cite{Lin:2021bcp} and ultimately of the Stolz--Teichner conjecture \cite{Segal,ST1,ST2}, for a large class of holomorphic CFTs.

\section{Generalizations}
\label{Sec:Generalizations}

\subsection{Non-abelian orbifolds and Generalized Moonshine}
\label{Sec:NonAbelian}

To construct $c=48$ holomorphic CFTs with few spin-two states (close to extremity), Gaiotto and Yin \cite{GaiottoYin} computed the (diagonal) orbifolds of the symmetric Monster CFT $\M^{\otimes 2} / \S_2$ by the following non-abelian subgroups of the Monster $\Mon$:\footnote{
    The (possible) class fusions for these subgroups were given in \cite{norton2002anatomy}.  $L_2(71)$ was proven to be a subgroup of $\Mon$ by \cite{holmes2008subgroups}.
}
\ie\label{NonAbelianSubgroups}
71:35, ~~ 59:29, ~~ 47:23, ~~ 29:7, ~~ L_2(29), ~~ \mathrm{A}_5, ~~ L_2(71).
\fe
Here, $p : n$ with $p$ prime and $n \dv p-1$ denotes the semidirect product $\bZ_p \rtimes \bZ_n$ under the natural group homomorphism $\bZ_n \to \mathrm{Aut}(\bZ_p) \cong \bZ_{p-1}$, $L_p(n)$ denotes the linear group over the finite field $\mathbb{F}_p$, and $\mathrm{A}_n$ is the alternating group on $n$ elements.
The numbers of spin-two states found under these orbifolds are summarized in Table~\ref{Tab:Xi}.\footnote{
    For $G = L_2(29), ~ \mathrm{A}_5, ~ L_2(71)$, when computing the twisted sector of an order-2 element $h$, Gaiotto and Yin \cite{GaiottoYin} only performed the partial projection by the cyclic subgroups $\vev{h} \cong \bZ_{14}, ~ \bZ_2, ~ \bZ_{36}$ of the full centralizers $\D_{28}, ~ \bZ_2 \times \bZ_2, ~ \D_{72}$, and obtained $Z[\M^{\otimes 2}/\S_2 \times G]_{q^0} = 222, 3816, 112$, respectively.  Using Norton's Generalized Moonshine \cite{norton1987generalized} data, we are able to uniquely determine the correct answers for $L_2(29)$ and $L_2(71)$, but not for $\mathrm{A}_5$ due to the author's lack of knowledge of the class fusion(s) for $\mathrm{A}_5 < \cen_\Mon(h)$.
}
They are all divisible by 12, even though the divisibility property \eqref{Divisibility} only requires them to be divisible by 6.

Orbifolds by general non-abelian subgroups of $\Mon$ present new challenges.  Previously, for cyclic orbifolds, the knowledge of the standard McKay--Thompson series (without spatial twist) was sufficient, because all involved twisted partition functions lie in their modular orbits.  The same is true for orbifolds by non-abelian subgroups in which the centralizer of every nontrivial element is a cyclic group; examples include the first four semidirect products in \eqref{NonAbelianSubgroups}.
However, general non-abelian orbifolds do not afford such convenience and require the data of Norton's Generalized Moonshine conjecture \cite{norton1987generalized}, which has been proven by Carnahan \cite{carnahan2010generalized,Carnahan:2009mjm,Carnahan:2012wk}.
Among the 194 conjugacy classes of $\Mon$, the generalized McKay--Thompson series for 137 have been computed by Norton, but 57 remain incomplete.\footnote{
    The author thanks Scott Carnahan for correspondence and for sharing Norton's data.
}

\begin{table}[h]
    \centering
    \begin{tabular}{|c|c|}
        \hline
        $G < \Mon$ & $Z[\M^{\otimes 2}/\S_2 \times G]_{q^0}$
        \\\hline
        $71:35$ & 156
        \\
        $59:29$ & 204
        \\
        $47:23$ & 276 
        \\
        $29:7$ & 1068
        \\
        $L_2(29)$ & 204
        \\
        $L_2(71)$ & 108
        \\\hline
    \end{tabular}
    \caption{Number of spin-two states (including the stress tensor) in the non-abelian orbifolds of the $c=48$ symmetric Monster $\M^{\otimes 2}/\S_2$ considered by Gaiotto and Yin \cite{GaiottoYin}.}
    \label{Tab:Xi}
\end{table}

To illustrate the nature of the non-abelian orbifold computation, and to explain the usage of Generalized Moonshine \cite{norton1987generalized}, we compute the orbifold of $\M^{\otimes 2}/\S_2$ by $L_2(71)$.
The orders, multiplicities and centralizers of the conjugacy classes of $L_2(71)$ can be obtained in any standard computer algebra system such as GAP \cite{GAP4} and are summarized in Table~\ref{Tab:L271}.  The class fusion for $L_2(71) < \Mon$ involves \cite{norton2002anatomy}
\ie\label{FusionL271}
    \text{1A, 2B, 3B, 4C, 5B, 6E, 7B, 9B, 12I, 18D, 35B, 36D, 71AB}.
\fe

To compute the projection of the untwisted sector by $L_2(71)$, we average $Z[\M^{\otimes 2}/S_2]^g$ over all $g \in L_2(71)$ and find 9 spin-two states.
For the twisted sectors, all conjugacy classes except for order-2 have centralizers that are cyclic groups, and hence, the projection can be easily achieved by performing modular transformations on the standard McKay--Thompson series.  The results are summarized in Table~\ref{Tab:L271}.

The remaining task is to compute the twisted sector of an order-2 element $h$ by performing the projection by $\D_{72}$. 
Let $\sigma$ denote the generator of the exchange $\S_2$.  The $h$-twisted sector of $(\M^{\otimes 2}/\S_2) / L_2(71)$ is the union of the $h$- and $h\sigma$-twisted sectors of $\M^{\otimes 2}/(\S_2 \times L_2(71))$.  It is clear that the $h\sigma$-twisted sector has no spin-two state, since
\ie 
    Z[\M^{\otimes 2}]_{h\sigma}(\tau) = Z[\M^{\otimes 2}]^{h\sigma}(-\frac{1}{\tau}) = Z[\M](-\frac{2}{\tau}) = Z[\M](\frac\tau2) = J(\frac\tau2)
\fe
has a vanishing constant Fourier coefficient.  The $h$-twisted sector of $\M^{\otimes 2}/(\S_2 \times L_2(71))$ is given by first taking the symmetric tensor product of the spin-one states in the $h$-twisted defect Hilbert space of the single-copy $\M$ and then projecting by the diagonal centralizer $\cen_{L_2(71)}(h) \cong \D_{72}$ to the trivial representation.

Consider the single-copy theory $\M$.  Let $g$ be an order-36 element such that $g^{18} = h$.  By performing modular transformations on the McKay--Thompson series for the classes \eqref{FusionL271} whose orders divide 36, i.e.\ 2B, 3B, 4C, 6E, 9B, 12I, 18D, 36D, we can compute $Z[\Mon]_h^{g^n}$ for all $n$ and thereby extract the $g$-eigenvalues of the 24 spin-2 states in the $h$-twisted defect Hilbert space.  The result is
\ie\label{EigVals}
    \{ -1_{\times 2}, \pm \omega, \pm \omega^2 \} \cup \{ \zeta^m, -\zeta^m_{\times 2} \mid m=1,2,4,5,7,8 \}, \quad \omega = e^{\frac{2\pi i}{3}}, \quad \zeta = e^{\frac{2\pi i}{9}}.
\fe 
These 24 states form 13 irreducible representations of $\D_{72}$.  Among them, 11 two-dimensional representations incorporating the complex $g$-eigenvalues can be unambiguously identified.  These two-dimensional representations, which we denote by $\mathbf{2}^i$ for $i=1,2,\dotsc,11$, all have vanishing trace under any reflection $r \in \D_{72}$.

\begin{table}[t]
    \centering
    \begin{tabular}{|c|c|c|c|}
        \hline
        Order & Multiplicity & Centralizer & Number of spin-two states
        \\\hline
        1 & 1 & $L_2(71)$ & 9
        \\
        2 & 1 & $\D_{72}$ & 16
        \\
        3 & 1 & $\bZ_{36}$ & 7 
        \\
        4 & 1 & $\bZ_{36}$ & 4 
        \\
        5 & 2 & $\bZ_{35}$ & 3 
        \\
        6 & 1 & $\bZ_{36}$ & 3 
        \\
        7 & 3 & $\bZ_{35}$ & 2 
        \\
        9 & 3 & $\bZ_{36}$ & 2 
        \\
        12 & 2 & $\bZ_{36}$ & 2 
        \\
        18 & 3 & $\bZ_{36}$ & 3 
        \\
        35 & 12 & $\bZ_{35}$ & 2
        \\
        36 & 6 & $\bZ_{36}$ & 2
        \\
        71 & 2 & $\bZ_{71}$ & 1
        \\\hline
    \end{tabular}
    \caption{The conjugacy classes of $L_2(71)$ organized by their orders, multiplicities, centralizers, and the numbers of spin-two states in the twisted sectors after projection by the centralizers.}
    \label{Tab:L271}
\end{table}

It is trickier to identify the representations of $\D_{72}$ for the $g$-eigenvalues $\{ -1_{\times 2} \}$ in \eqref{EigVals}, since there are two distinct one-dimensional representations with $g$-eigenvalue $-1$.  These two representations differ by their reflection $r$-eigenvalues $\pm 1$ and are denoted correspondingly by $\mathbf{1}^\pm$.  There are three possibilities: $\{ \mathbf{1}^\pm \}$, $\{ \mathbf{1}^+_{\times 2} \}$ and $\{ \mathbf{1}^-_{\times 2} \}$.  To fix this, we use the following piece of Norton's data of Generalized Moonshine \cite{norton1987generalized}.  
Let $g$ be an order-2 element of the centralizer of $h$.
According to Norton, the possible constant Fourier coefficients of $Z[\M]_h^g$ are $0, \pm 8, \pm 24$, but not $\pm 2$.  This means that the $g$-eigenvalues $\{ -1_{\times 2} \}$ must belong to two \emph{different} one-dimensional representations of $\D_{72}$, namely $\{ \mathbf{1}^\pm \}$, such that $Z[\M]_h^r$ has vanishing constant Fourier coefficient.

The direct sum of the $\D_{72}$ representations of the spin-one states in the single-copy $\M$ is $\mathbf{r} = \mathbf{1}^+ \oplus \mathbf{1}^- \oplus_{i=1}^{11} \mathbf{2}^i$.  To count the number of spin-two states in the $h$-twisted sector of $(\M^{\otimes 2}/\S_2) / L_2(71)$, we take the symmetric tensor product of $\mathbf{r}$ with itself and count the number of trivial representations, finding 16.
The total number of spin-two states in $\M^{\otimes 2}/\S_2 \times L_2(71)$ is thus 108, which is divisible by 12 and in particular satisfies the divisibility property \eqref{Divisibility}.  By contrast, had we assumed that the $g$-eigenvalues $\{ -1_{\times 2} \}$ in \eqref{EigVals} belonged to the same representation of $\D_{72}$, namely $\{ \mathbf{1}^+_{\times 2} \}$ or $\{ \mathbf{1}^-_{\times 2} \}$, then we would have concluded that there were 17 spin-two states in the $h$-twisted sector, giving a total of 109 spin-two states in $\M^{\otimes 2}/\S_2 \times L_2(71)$, and violating the divisibility property \eqref{Divisibility}.

\subsection{Symmetric orbifolds of Schellekens' list}
\label{Sec:Schellekens}

The set of all $c=24$ bosonic holomorphic CFTs (with a unique vacuum) was conjectured by Schellekens \cite{Schellekens:1992db} (see also \cite{vanEkeren:2017scl}), and proven in \cite{Moller:2019tlx,Hohn:2020xfe,vanEkeren:2020rnz} except for the uniqueness of the Moonshine Module.
We denote them by $\Sch^\ell$, where
\ie 
    Z[\Sch^\ell](\tau) = J(\tau) + 12\ell,
\fe
with $\ell$ taking values in \eqref{Schellekens}.  As explained in {Section}~\ref{Sec:Arf}, without demanding a unique vacuum, $\ell$ can in fact be any integer.
We denote by $\Sch^\ell$ \emph{any} such direct sum theory, and by $\tilde\Sch^\ell$ the tensor product of $\Sch^\ell$ with the Kitaev chain \cite{Kitaev:2000nmw}/invertible field theory $(-1)^\text{Arf}$.  
This subsection examines the topological modularity of their permutation orbifolds, i.e.\ the satisfaction of the divisibility property \eqref{Divisibility}.

The DMVV formula \eqref{DMVV0} applied to $\tilde\Sch^\ell$ or $\Sch^\ell$ is related to that applied to $\tilde\M$ or $\M$ by a simple multiplicative factor,
\ie\label{SchEG}
    \cZ_\text{ell}[\tilde\Sch^\ell](\sigma,\tau) = \cZ[\Sch^\ell](\sigma,\tau)^{-1}
    = p (j(\sigma) - j(\tau)) \left( \frac{\Delta(\sigma)}{p} \right)^{\frac\ell2},
\fe 
where $\Delta(\sigma) = \eta(\sigma)^{24}$ and $\eta$ is the Dedekind eta function.
The generating functions for the constant Fourier coefficients are likewise related,
\ie 
    \cZ_\text{ell}^\text{const}[\tilde\Sch^\ell](\sigma) &= \cZ_\text{ell}[\tilde\M](\sigma) \left( \frac{\Delta(\sigma)}{p} \right)^{\frac\ell2} = p J(\sigma) \left( \frac{\Delta(\sigma)}{p} \right)^{\frac\ell2},
    \\
    \cZ^\text{const}[\Sch^\ell](\sigma) &= \cZ^\text{const}[\Sch](\sigma) \left( \frac{\Delta(\sigma)}{p} \right)^{-\frac\ell2} = - p^{-2} \left( \frac{dJ(\sigma)}{dp} \right)^{-1} \left( \frac{\Delta(\sigma)}{p} \right)^{-\frac\ell2}.
\fe

Both expressions are divisible with base 12 for any $\ell$, which can be seen as follows.  First,
\[
    pJ(\sigma), \quad -p^{-2} \left( \frac{dJ(\sigma)}{dp} \right)^{-1}
\]
are divisible with base 24 by Proposition~\ref{Main}.  Next,
\[
    \left( \frac{\Delta(\sigma)}{p} \right)^{-\frac\ell2}
\]
is divisible with base $12\ell$ by Corollary~\ref{Eta}.
Therefore, their product is divisible with base 12 according to Lemma~\ref{Closure}.  We have thus proven the divisibility property \eqref{Divisibility} for all symmetric orbifolds of all $c=24$ bosonic holomorphic CFTs possibly coupled to $(-1)^\text{Arf}$.

\subsection{Prospects}

A natural generalization of this work is to consider the Conway CFT (Super-Moonshine Module) constructed by Duncan \cite{duncan2007super}, which is a $c=12$ supersymmetric holomorphic CFT realizing the Conway group $\text{Co}_1$ as its global symmetry.\footnote{
    See \cite{Gaiotto:2018ypj,Johnson-Freyd:2020wh,Theo} for studies of the Conway CFT \cite{duncan2007super} in the context of TMF.
}
The time-twisted elliptic genera are constants given by the characters of the 24-dimensional representation of $\text{Co}_1$ \cite{duncan2015moonshine}.  This constancy allows one to compute the elliptic genera of various orbifolds by non-anomalous subgroups with greater ease, and prove divisibility \cite{Albert:2022gcs}.

The last two subsections have obvious extensions.
One can consider additional non-abelian subgroups of Monster beyond those discussed in {Section}~\ref{Sec:NonAbelian}, and also study more than two copies.  Depending on the subgroup, the available Generalized Moonshine \cite{norton1987generalized} data may or may not be enough to fix the answer.  It would be interesting if the divisibility property could help determine some currently unknown generalized McKay--Thompson series.  
Concerning {Section}~\ref{Sec:Schellekens}, we note that other theories in Schellekens' list \cite{Schellekens:1992db} also boast global symmetries to orbifold.

The fact that the symmetric orbifolds of the Monster CFT \cite{frenkel1984natural,frenkel1988vertex} coupled to the Kitaev chain \cite{Kitaev:2000nmw} have constant elliptic genera seems intriguing.   
There is clearly no $(1,1)$ supersymmetry since the Neveu--Schwarz sector is identical to the Monster CFT and only contains operators of integer spin.\footnote{
    The author thanks Theo Johnson-Freyd for a discussion.
}
By \eqref{SchEG}, the constancy of the elliptic genera is lost upon replacing the Monster CFT by any other $c=24$ holomorphic CFT.

Some observed divisibility---or, topological modularity---in this note can be proven in a number-theoretic fashion, after leveraging the DMVV formula \cite{Dijkgraaf:1996xw} and the twisted denominator formula \cite{norton1984more,borcherds1992monstrous}.
Note that the series divisibility of \emph{pure} permutation orbifolds, which in the Monster context means the series divisibility of $J$, follows from Ganter's work \cite{ganter2006orbifold} on power operations.\footnote{
    The series divisibility of $pJ$ with base \emph{24} instead of 12 can be explained by considering symmetric orbifolds of the TMF class $j \Delta^{-1}$, instead of $j \Delta^{-2}$ that is realized by Monster. This comment is due to Theo Johnson-Freyd.
}
In future, we may hope for a Monster-equivariant TMF class, together with a fully developed theory, extending \cite{Barthel:2020fva}, of power operations on equivariant TMF.

\section*{Acknowledgements}

The author thanks Jan Albert, Scott Carnahan, Chi-Ming Chang, Theo Johnson-Freyd, Justin Kaidi, and Yifan Wang for discussions and comments on the draft.
This work was supported by the Simons Collaboration Grant on the Non-Perturbative Bootstrap.



%

\bibliography{refs}
\bibliographystyle{JHEP}

\end{document}